\documentclass[amssymb,aps,twocolumn,showpacs,nofootinbib]{revtex4}
\usepackage[]{graphicx}
\def\ket#1{| #1 \rangle}
\def\bra#1{\langle #1 |}
\def\bracket#1#2{\langle #1 | #2 \rangle}

\begin{document}

\title{Higher Security Thresholds for Quantum Key Distribution by Improved Analysis of Dark Counts}

\author{J.-C. Boileau$^{1,2}$, J. Batuwantudawe$^1$, R. Laflamme$^{1,2}$}

\affiliation{$^1$Institute for Quantum Computing, University of Waterloo, Waterloo, ON, N2L 3G1, Canada.\\
$^2$Perimeter Institute for Theoretical Physics, 35 King Street North, Waterloo, ON, N2J 2W9, Canada.}

\date{\today}

\begin{abstract}

We discuss the potential of quantum key distribution (QKD) for long distance communication by proposing a new analysis of the errors caused by dark counts. We give sufficient conditions for a considerable improvement of the key generation rates and the security thresholds of well-known QKD protocols such as Bennett-Brassard 1984, Phoenix-Barnett-Chefles 2000, and the six-state protocol. This analysis is applicable to other QKD protocols like Bennett 1992. We examine two scenarios: a sender using a perfect single-photon source and a sender using a Poissonian source.
 
\end{abstract}


\maketitle

The goal of quantum key distribution (QKD) is to extend a shared secret key for use as a one-time pad to encode classical messages. The advantage of QKD is that its security is based on the laws of quantum mechanics and not on the unproven complexity of a mathematical problem as in classical cryptography. These last few years, many encouraging experiments demonstrated QKD, some spanning more than a hundred kilometers through optical fibers~\cite{GYS04}. The main source of errors is usually due to dark counts from the detectors. A dark count is when a detector fires independently (or in the absence) of a qubit state encoded by the sender, Alice. If qubit losses are considerable, then the receiver, Bob, will receive many empty pulses, and dark counts from his detectors will induce a high error rate.

In this paper, for simplicity, we refer specifically only to four different QKD protocols: Bennett 1992 (B92), Phoenix-Barnett-Chefles 2000 (PBC00), Bennett-Brassard 1984 (BB84), and the six-state protocol, which are two, three, four, and six state protocols, respectively~\cite{B92, PBC00, BB84, B98}. In B92, Alice encodes random bits using two non-orthogonal states, say $\ket{\psi_1}$ and $\ket{\psi_2}$, and sends them to Bob. He makes the measurement corresponding to the Positive Operator-Valued Measure (POVM) $\{\alpha\ket{\overline{\psi}_1}\bra{\overline{\psi}_1}, \alpha\ket{\overline{\psi}_2}\bra{\overline{\psi}_2}, \openone - \alpha\ket{\overline{\psi}_1}\bra{\overline{\psi}_1}-\alpha\ket{\overline{\psi}_2}\bra{\overline{\psi}_2} \}$, where $\ket{\overline{\psi}_j}$ is orthogonal to $\ket{\psi_j}$ and $\alpha$ equals $\frac{1}{1+|\bracket{\psi_1}{\psi_2}|}$. Bob's measurement either determines which state Alice did not send (from which Bob can deduce the encoded bit) or is inconclusive. PBC00 is similar to B92 but uses three non-orthogonal states, say $\ket{\psi_1}$, $\ket{\psi_2}$ and $\ket{\psi_3}$, that form an equilateral triangle in the X-Z plane of the Bloch sphere. She encodes her random bits using random bases from either $\{\ket{\psi_1}, \ket{\psi_2}\}$, $\{\ket{\psi_2}, \ket{\psi_3}\}$, or $\{\ket{\psi_3}, \ket{\psi_1}\}$. Bob performs the POVM $\{\frac{2}{3}\ket{\overline{\psi}_1}\bra{\overline{\psi}_1}, \frac{2}{3}\ket{\overline{\psi}_2}\bra{\overline{\psi}_2}, \frac{2}{3}\ket{\overline{\psi}_3}\bra{\overline{\psi}_3}\}$. After Bob measures all of the qubits, Alice declares publicly which basis she used for each. By deduction, Bob can sometimes retrieve Alice's state. Alice and Bob discard the other results. It can be shown that, neglecting the qubit losses, the rate of conclusive results is $\frac{1}{2-e_x}$ where $e_x$ is the bit error rate. A {\it conclusive} result corresponds to any pair of qubits not discarded by Alice and Bob. 

To implement BB84, Alice encodes a random bit in either $\{\ket{0}, \ket{1}\}$ or its conjugate basis $\{\ket{+}, \ket{-}\}$. For each qubit, Bob randomly measures in one of these bases. They only keep results for which they used the same basis. The six-state protocol is identical to BB84 except that Alice and Bob choose from three different bases: $\{\ket{0}, \ket{1}\}$,$\{\ket{+}, \ket{-}\}$, and $\{\frac{1}{\sqrt 2}(\ket{0}+i\ket{1}),\frac{1}{\sqrt 2}(\ket{0}-i\ket{1})\}$. We can modify BB84 and the six-state protocol by choosing bases with non-equal probabilities, increasing the chance of agreement~\cite{LCA04}. The rate of results for which identical bases are used converges asymptotically to 1. Below, we calculate the key generation rates of BB84 and the six-state protocol using this asymptotic result.

Mayers~\cite{M96} produced the first unconditional security proof of BB84. Shor and Preskill~\cite{SP00} proposed a simpler proof based on ideas from Lo and Chau~\cite{LC99}. Their security proof has been generalized to other protocols including B92, PBC00, and the six-state protocol~\cite{TKI03TL04, BTBLR05, L01, RG05}. We improve the secret key generation rate of these QKD protocols by proposing a slight modification of these proofs. Our main idea is based on a variation of a theorem proved in Ref.~\cite{GLLP02}. We assume that an eavesdropper, Eve, can perform any attack consistent with quantum mechanics, but cannot get any information about Alice's or Bob's labs or control their apparatus. We discuss later how realistic these assumptions are and how it is possible to slightly relax them. We study two cases: one where Alice's source can create a single photon on demand, and another where it follows a Poisson distribution. For simplicity, we give details only about Shor and Preskill's security proof of BB84 and not other protocols.

At the end of this paper, we compare the updated error rate thresholds and key generation rates of BB84, PBC00, and the six-state protocol with previous results. The same arguments could improve other QKD protocols, including B92. However, B92's phase estimation bound depends on qubit losses in the channel and the number of inconclusive results, complicating the analysis. Since our goal is to describe a general technique to improve security thresholds, we only treat the simpler cases as examples.

The Shor and Preskill proof first shows the security of an entanglement distillation protocol (EDP) for QKD, and subsequently reduces the EDP to BB84. For convenience, we define $\ket{\Phi^{\pm} }=\frac{1}{\sqrt2}(\ket{0}\ket{0}\pm\ket{1}\ket{1})$ and $\ket{\Psi^{\pm} }= \frac{1}{\sqrt2}(\ket{0}\ket{1}\pm\ket{1}\ket{0})$.

The structure of the {\bf EDP that can be reduced to BB84} in Shor and Preskill's proof is as follows:

\noindent{\bf 1.} Alice creates $n$ pairs of the form $\ket{\Phi^+}$ and sends the second half of each pair to Bob after randomly applying the identity or the Hadamard gate on it.

\noindent{\bf 2.} After Bob confirms that he has received all of Alice's states, Alice publicly declares the random rotation that she used on each qubit. Bob undoes the transformations on the corresponding qubits.

\noindent{\bf 3.} With no eavesdropping or channel noise, Alice and Bob will share $n$ perfect pairs of the form $\ket{\Phi^+}$. They can now measure their qubits in the same basis to share a secret key. However, noise and eavesdropping induce errors. If the bit and the phase error rates are low enough, then error correction can be applied to obtain $m$ perfect pairs of the form $\ket{\Phi^+}$ where $m\leqslant n$.

\noindent{\bf 4.} Alice and Bob can estimate the bit error rate by comparing bit measurements from a sample of pairs, called test bits. A bit (or X) error on a pair occurs when Alice and Bob share either $\ket{\Psi^+}$ or $\ket{\Psi^-}$. A phase (or Z) error corresponds to $\ket{\Phi^-}$ or $\ket{\Psi^-}$. A Y error corresponds to $\ket{\Phi^-}$ or $\ket{\Psi^+}$. Y error estimation could provide information about the correlation between bit and phase errors. Because Alice randomly applies the identity or Hadamard gate, it can be shown that the bit error rate, $e_{x}$, and the phase error rate, $e_{z}$, are approximately equal, independent of channel noise and Eve's strategy. In BB84, Alice and Bob have no information about Y errors.

\noindent{\bf 5.} Depending on the bit error rate measured on the test bits, Alice and Bob apply error correction on the other pairs. If we suppose one-way error correction using CSS codes~\cite{CSS}, a lower bound for generation rate $\frac{m}{n}$ for the perfect pairs is given asymptotically by
\begin{eqnarray}
S = p_{c}[1-H(e_x,e_z)]
\label{keyGen1}
\end{eqnarray}
where $H$ is the Shannon entropy ($H(e_x,e_z)= H(e_x)+H(e_z|e_x)$ is the entropy of the bit-phase error pattern) and $p_{c}$ is the rate of conclusive results. For simplicity, we assume that the proportion of test bits is negligible.
$\square$

Shor and Preskill showed that this EDP, and thus BB84, were unconditionally secure with a key generation rate given by Eq.~\ref{keyGen1}. Since $H(e_x)$ is asymptotically the fraction of bits sacrificed for bit error correction, it implies that $H(e_z|e_x)$ is an upper bound on the fraction of information that Eve has about the key after bit error correction. A consequence is that privacy amplification, as introduced in Ref.~\cite{BBR88}, can be used to simplify the post-processing of the QKD protocol. As shown in Ref.~\cite{RK04}, privacy amplification can generate a secret key by sacrificing a number of bits asymptotically proportional to Eve's information.

The reduction of the EDP to BB84 assumes that Alice uses a source which emits a single photon on demand. In a more realistic situation, Alice's source would emit a photon pulse following a Poisson distribution. Unfortunately, when Alice sends two or more photons containing the same quantum information at the same time, Eve can measure one to gain information about the key without detection.  Accounting for this attack (but assuming Eve has no information about the random phase of the signal emitted by a coherent light source), a more general equation of the secret key generation rate, combining results from Ref.~\cite{GLLP02} and Ref.~\cite{L01}, and using the improvement suggested in Ref.~\cite{L05}, is given asymptotically by
\begin{eqnarray}
S = p_{c}[\omega_0 + \omega_1-H(e_x)-\omega_1 H(e_z^{1}| e_x)]
\label{keyGen2}
\end{eqnarray}
where $\omega_1$ is the fraction of the conclusive results corresponding to single-photon pulses, $\omega_0$ is the fraction of the conclusive results corresponding to empty pulses (the presence of background noise, for example), and $e_x^{1}$ ($e_z^{1}$) is the bit (phase) error rate restricted to conclusive results from single-photon pulses. $e_x$ ($e_z$) is still the bit (phase) error rate over all conclusive results. If Alice has a source that emits a single photon on demand, then $\omega_0=0$, $\omega_1 = 1$, $e_j^{1}=e_j$ for $j \in \{x,y,z\}$, and $S=p_{c}[1- H(e_x, e_z)]$ as expected.

To prove Eq.~\ref{keyGen2}, it was argued that since Alice and Bob want an identical key and cannot differentiate multi-photon from single-photon pulses, they must correct all bit errors, asymptotically losing a fraction $H(e_x)$ of the results in the process. To apply privacy amplification on the remaining bits and obtain a secret key, Alice and Bob must upper bound Eve's information. If we assume that the phase of the signal is random\footnote{Recently, it was shown the Eve could use extra information about the phase of the signal to her advantage~\cite{LP05}, though the extent is unknown.}, there is no coherence between states with different photon numbers. Thus, we can categorize each bit of the resulting key as being associated with an empty, single-, or multi-photon pulse. Assuming the worst case, Eve has full information about the results associated with multi-photon pulses. On the other hand, she has no information about Alice's bits corresponding to empty pulses. By the Shor-Preskill's arguments discussed earlier, the fraction of information that Eve could extract from the results corresponding to single-photon pulses is upper bounded by $H(e_z^{1}|e_x)$. Consequently, Eve's information about Alice's remaining key is upper bounded by $(1-\omega_0-\omega_1)+\omega_1 H(e_z^{1}|e_x)$. After privacy amplification, Eve has no information about Alice's key. The same is true of Bob's key since it is identical to Alice's. Therefore, the secret key generation rate is given by Eq.~\ref{keyGen2}.

Similarly, since Shor-Preskill's proof can be adapted to B92, PBC00 and the six-state protocol~\cite{TKI03TL04, BTBLR05, L01}, these protocols can be shown unconditionally secure with a key generation rate given by Eq.~\ref{keyGen2}.

The above argument does not differentiate between a single photon emitted by Alice that is successfully measured by Bob and a single photon that is lost in the channel (or taken by Eve) followed by a dark count measured by Bob. However, these cases may be analyzed separately. Consider the following four types of conclusive results. 

\begin{enumerate}
\item Successful measurement of a {\it qubit state} (physically corresponding to a photon received from the channel) that originated from a single-photon pulse. Note that the qubit state could have been manipulated by Eve. 
\item Successful measurement of a qubit state that originated from a multi-photon pulse. 
\item Empty pulses from Alice followed by a successful measurement of a qubit state by Bob (ie. Eve may send a qubit state to Bob even if Alice emits nothing).
\item Dark count events:  Bob doesn't receive a qubit state, but one of his detectors fires. 
\end{enumerate}

The dark count events are independent of Alice's or Eve's actions. We define $p_{c}^{emp}$, $p_{c}^{sq}$, and $p_{c}^{mq}$ as the rate of conclusive results corresponding to qubit states, received by Bob, associated with empty pulses, single-photon pulses, and multi-photon pulses, respectively. We define $p_{c}^{dk}$ as the rate of conclusive results associated with dark counts. Note that 
\begin{equation}
p_{c}=p_{c}^{emp}+p_{c}^{sq}+p_{c}^{mq}+p_{c}^{dk}
\end{equation}

We remark that the background noise has two different contributions: intrinsic and extrinsic. The intrinsic contribution is caused by elements from Bob's lab while the extrinsic contribution is from external sources. The sun and backscattering light in two-way QKD are examples of external sources of background noise. Based on our assumptions, Eve may control the external sources of background noise, but not the ones inside Bob's lab. Following our previous definitions, the only contribution to $p_{c}^{dk}$ is intrinsic. Any external sources will contribute to $p_{c}^{emp}$, $p_{c}^{sq}$, and $p_{c}^{mq}$ since they correspond to Bob receiving a qubit state from the channel. For convenience, in this paper, {\it dark counts} always refer to the intrinsic contribution of background noise. We assume for simplicity that dark counts are independent of other measurement results.

We now explain how it is possible to achieve a better bound for the secret key generation rate than Eq.~\ref{keyGen2}. As before, a fraction $H(e_x)$ of the results are lost due to bit error correction. Assuming again that the phase of the signal is random from Eve's perspective, each bit of the resulting key corresponds to one of the four types of conclusive results described above. From previous arguments, Eve has a fraction $H(e_z^{sq}|e_x)$ of information about conclusive results from Category 1 and, in the worst case scenario, full information about those from Category 2. $e_x^{sq}$ and $e_z^{sq}$ are defined as the bit and phase error rates on the conclusive results restricted to Category 1. When Alice emits an empty pulse and it is followed by a successful measurement of a qubit state by Bob, we assume that the qubit state was created by Eve. A conservative assumption is that Eve has full information about Bob's results from Category 3.\footnote{In the case of B92, it is easy to show that this assumption is necessary, but it might be too strict for other protocols like PBC00, BB84, and the six-state protocol.} 
Supposing dark count rates are the same in all detectors and independent of Eve and other measurement results, Bob's results from Category 4 are completely random and Eve has no information about them
\footnote{For simplicity, we suppose that the dark count rates are uniform over all detectors and that they are independent of other measurement results. If dark count rates differ from detectors, we suggest two options. In one, Bob uses a random transformation to switch the role of the detectors in the measurement. For example, in BB84,  Bob could apply, at random, an extra $Y$ operation on the received qubits to switch the role of the detectors when measuring in the $\{\ket{0}, \ket{1}\}$ and $\{\ket{+}, \ket{-}\}$ bases. A second option is to bound Eve's information from an estimate of the probability that a detector fires relative to the others in the case of a dark count. Assuming dark counts are independent of other measurement results, in BB84 and the six-state protocol, with only two detectors, Eve's information is bounded by $1-H(q)$ where $q$ is the probability that the first detector fires in the case of a dark count. It is interesting to note that if Eve has some control over the probability $q$ and could change it from one dark count event to another, then, by entropic concavity, Eve's information is bounded by $1-H(q_{worst}^{ave})$, where $q_{worst}^{ave}$ is the worst estimate of the average of $q$. Determining the value of $q_{worst}^{ave}$ can be very hard, but it is related to the level of confidence that Alice and Bob have on their ability to counter or detect Eve if she tries to change the properties of the detectors. Similarly, if dark counts are correlated to other measurement results, we can upper bound Eve's information with restrictions on the correlations. }
. Consequently, the fraction of information that Eve has on Bob's key after bit error correction is upper bounded by $\frac{1}{p_{c}}(p_{c}^{emp}+p_{c}^{mq}+p_{c}^{sq}H(e_z^{sq}|e_x))$. Therefore, the secret key generation rate is lower bounded by 
\begin{eqnarray}
S_b = p_{c}^{sq}+p_{c}^{dk}-p_{c}H(e_x)-p_{c}^{sq}H(e_z^{sq}|e_x).
\label{keyGen3}
\end{eqnarray}
We emphasize that it is not necessary for Alice and Bob to know which events correspond to each class of conclusive results.

In the derivation of Eq. ~\ref{keyGen3}, we bounded Eve's information about Bob's key. However, we could have instead bounded Eve's information about Alice's key. In this case, Eve has no information about the bit chosen by Alice when she sends a vacuum states. But she could have some information about Alice's portion of the key corresponding to dark counts (unless Alice sent an empty pulse). Using similar arguments, we obtain 
\begin{eqnarray}
S_a = p_{c}^{sq}+p_{c}\omega_0-p_{c}H(e_x)-p_{c}^{sq}H(e_z^{sq}|e_x).
\label{keyGen4}
\end{eqnarray}

Combining Eq.~\ref{keyGen3} and Eq.~\ref{keyGen4}, we obtain a new lower bound for the secret key generation rate,

\begin{eqnarray}
S&=& \max[S_a, S_b].
\label{keyGen5}
\end{eqnarray}

Remark that the concavity of entropy and $\omega_1e_z^{1}= \frac{p^{sq}}{p^{c}}e_z^{sq}+(\omega_1-\frac{p^{sq}}{p^{c}})e_z^{dk}$ imply that $\omega_1H(e_z^{1}| e_x))\geqslant \frac{p^{sq}}{p^{c}}H(e_z^{sq}| e_x)+(\omega_1-\frac{p^{sq}}{p^{c}})H(e_z^{dk}| e_x)$. We can rewrite this as $\omega_1(1-H(e_z^{1}| e_x)) \leqslant  \frac{p^{sq}}{p^{c}}(1-H(e_z^{sq}| e_x))$, since it can be argued that $e_z^{dk}=\frac{1}{2}$. Therefore, the secret key generation rate given by Eq.~\ref{keyGen4} (and Eq.~\ref{keyGen5}) is always greater than or equal to the one given by Eq.~\ref{keyGen2}.

To evaluate Eq.~\ref{keyGen5}, Alice and Bob must be able to determine all quantities involved in it. For this purpose, we study two different situations: Alice has a source that emits a single photon on demand or one that follows a Poisson distribution.

In both situations, $e_x$ is estimated from test bits, and $p_{c}^{dk}$ can be calculated from the predetermined dark count probability $C$ of the detectors and the number of empty pulses not associated with dark counts that Bob receives. If $C$ is not fixed, Bob might block his detection units randomly and estimate $p_{c}^{dk}$ from these results. For this to be true, it is important that Eve is not allowed to reduce the dark count probability without being detected. But is this a valid assumption? In practice, Eve could try to cool down the detectors or send bright pulses to disable them at will. Furthermore, there might be some uncertainty in the measurement of $p_{c}^{dk}$, even in the absence of an eavesdropper. Since a dark count could be interpreted as Eve sending a random state to Bob, we remark that lower bounds for $C$ and $p_{c}^{dk}$ are sufficient to obtain a better key generation rate using Eq.~\ref{keyGen5}. Establishing a high level of confidence on a lower bound for $p_{c}^{dk}$ seems very hard in practice. However, it might be possible through experimental research and tests on reducing dark count rates of detectors.

If Alice has a source that emits single photons, $\omega_0=0$ and $p_{c}^{mq}=0$, then Eq.~\ref{keyGen5} reduces to Eq.~\ref{keyGen3} and $e_x= \frac{1}{p_{c}}(p_{c}^{sq}e_x^{sq}+p_{c}^{dk}e_x^{dk})$, where $e_x^{dk}$ is the bit error rate over conclusive events associated with dark counts. $e_x^{dk}=\frac{1}{2}$ which implies that Bob can estimate $e_x^{sq}$ from the value of $e_x$ measured on test bits. $H(e_z^{sq}|e_x)=H(e_z^{sq}|e_x^{sq})$ can be evaluated depending on the protocol used. It can easily be shown that, for the six-state protocol, $e_x^{sq}=e_y^{sq}=e_z^{sq}$~\cite{L01}. For BB84,  $e_x^{sq}=e_z^{sq}$ and $0\leqslant e_y^{sq}\leqslant 2e_x^{sq}$~\cite{SP00}. For PBC00, it was shown that  $e_z^{sq}=\frac{5}{4}e_x^{sq}$ and $\frac{1}{4}e_x^{sq}\leqslant e_y^{sq}\leqslant \frac{9}{4}e_x^{sq}$~\cite{BTBLR05}.

In the absence of errors due to dark counts, $p_{c}^{dk}=0$. By solving $S(e_x)=0$, we find that the bit error rate threshold is 12.6\% for the six-state protocol, 11.0\% for BB84, and 9.81\% for PBC00. If we now suppose that $e_x^{sq}$ is fixed, then the bit error rate threshold increases as shown in Tab.~\ref{table1}. Note that the bit error rate threshold depends on the contribution of errors not associated to dark counts.

\begin{table}[htdp]
\caption{Bit error rate thresholds for BB84, PBC00, and the six-state protocol using a single-photon source and assuming fixed values of $e_x^{sq}$, which is the bit error rate of the results not associated with dark counts.}
\begin{center}

\begin{tabular}{|c||c|c|c|}
\hline
&$e_x^{sq}=0$&$e_x^{sq}=0.01$&$e_x^{sq}=0.1$\\ \hline \hline
{\rm PBC00} & 50\% & 43\% & -  \\ \hline
{\rm BB84} & 50\% & 44\% & 13\% \\ \hline
{\rm Six-State Protocol}& 50\% & 46\% & 19\% \\ \hline

\end{tabular}
\end{center}
\label{table1}
\end{table}

Tab.~\ref{table1} reflects the potential of a special analysis for dark counts.  For any of the previous QKD protocols, if the errors are only caused by dark counts ($e_x^{sq}=0$), then the bit error rate threshold is $\frac{1}{2}$, which implies there is no bound on the distance for communication. However, we must keep in mind that this result is derived using many special conditions. In practice, $e_x^{sq}$ is non-zero and, since there is decoherence in the channel and extrinsic sources of background noise, $e_x^{sq}$ usually increases with the distance of communication. We also assumed that Alice and Bob perfectly know the dark counts rates of their detectors, that they are the same for all detectors, that they are independent of other measurements, and that Eve cannot lower them. However, even if one or more of these assumptions are not respected, it is still possible to slightly modify Eq.~\ref{keyGen5}, as we explained earlier, and obtain an improvement over Eq.~\ref{keyGen2}.

\begin{figure}[hb!]
 \includegraphics[scale=0.5]{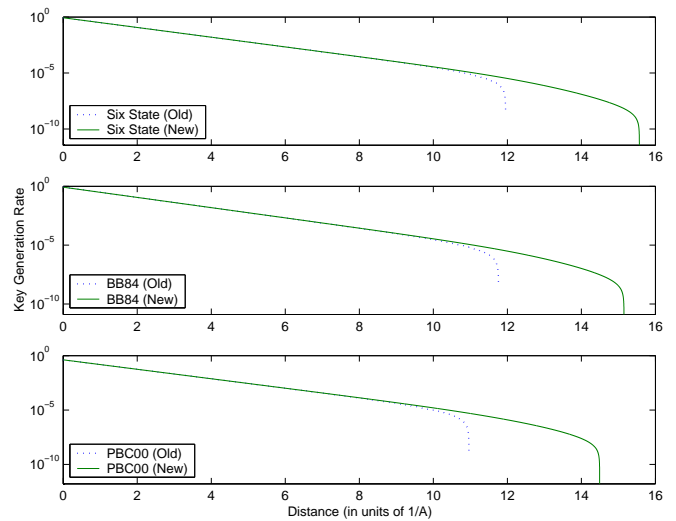}
\caption{\footnotesize{Semi-log graph of the key generation rate of PBC00, BB84, and the six-state protocol as a function of distance, $l$, for $e_x^{sq}=0.01$ and $C=10^{-6}$ calculated using the old method (Eq.~\ref{keyGen2}) and the new one (Eq.~\ref{keyGen5}) assuming a perfect single-photon source ($p_{c}^{mq}=0$).}}
\label{Fig1}
\end{figure}

In Fig.~\ref{Fig1}, we observe that the new method of calculating the key generation rate, using Eq.~\ref{keyGen5}, improves the achievable distance for PBC00, BB84, and the six-state protocol assuming a single-photon source. For simplicity, we suppose that the dark count probability, $C$, is the same for all detectors and that $e_x^{sq}$ is fixed and independent of distance. We assume no qubit losses at $l=0$, where $l$ is the length of the channel, and neglect events when two different detectors fire simultaneously. Under these conditions, for BB84 and the six-state protocol, $p_{c}^{sq}\approx \eta$ and $p_{c}^{dk}\approx 2C(1-\eta)$, where $\eta=e^{-Al}$ is the probability that a photon successfully travels through the channel and $A$ is the attenuation in the fiber. For PBC00, $p_{c}^{sq}\approx \frac{1}{2-e_x}\eta$ and $p_{c}^{dk}\approx 2C(1-\eta)$. Note that, since $\frac{p_{c}^{dk}}{p_{c}}$ is always equal or higher in PBC00 than for BB84 or the six-state protocol, PBC00's maximum achievable distance is lower for the same bit error rate.

We now consider the case where Alice uses a source that follows a Poisson distribution ($p_{c}^{mq}\neq0$). We only provide the result for BB84, but our arguments are valid for other QKD protocols, including B92, PBC00, and the six-state protocol.

\begin{figure}[hb!]
 \includegraphics[scale=0.5]{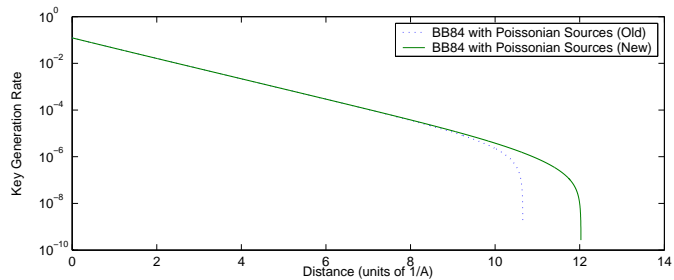}
\caption{\footnotesize{Semi-log graph of the key generation rate of BB84 as a function of distance, $l$, for $e_x^{sq}=0.01$ and $C=10^{-6}$ assuming a Poissonian source and combined with the decoy state method with $\bar{\mu}=0.5$. We compare the key generation rates calculated using Eq.~\ref{keyGen2} and Eq.~\ref{keyGen5}.}}
\label{Fig3}
\end{figure}

Decoy states~\cite{H03} could be used to evaluate $p_{c}^{sq}$ and $e_x^{sq}$ precisely. Ref.~\cite{W04, LMC04} explain how Alice could randomly vary the average photon number, $\mu$, of her source to obtain, from statistics, precise estimates of the rate of conclusive results associated with single-photon pulses, $p_{c}\omega$, and the corresponding bit error rate, $e_x^{1}$. $p_{c}^{sq}$ and $e_x^{sq}$ can be easily derived from the following two relations: $p_{c}\omega=p_{c}^{sq}+2Ce^{-\bar{\mu}}\bar{\mu}(1- \eta)$ and $e_x^{1}=e^{-\bar{\mu}}\bar{\mu} (\eta e_x^{sq}+2C(1-\eta)e_x^{dk})/(p_{c}\omega)$, where $\bar{\mu}$ is the global average photon number. Fig.~\ref{Fig3} shows that the decoy state method can also be improved by using Eq.~\ref{keyGen5}.

If we don't use decoy states, a worst case estimate of $p_{c}^{sq}$ and $e_x^{sq}$ is possible. However, Eq.~\ref{keyGen5} provides only a small improvement since, without decoy states, multi-photon pulses are usually a much more important limiting factor than dark counts.

In this paper, we showed that a high confidence in the stability of the dark counts of the detectors against the possible attack of an eavesdropper implies a significant increase of the robustness of most QKD protocols against dark counts, one of most important contributors of noise in quantum communication. We studied particularly the cases of PBC00, BB84 and the six-state protocol. We explained how to get an improvement of the secret key generation rate and of the achievable distance in some non-ideal situations, including when Alice uses a Poissonian photon source, when Alice and Bob know only a lower bound for the dark count rates of their detectors, and when the dark count rates are not uniform over the detectors. Further improvements to the secret key generation rate might come from using two-way error correction~\cite{GL01} and by artificially adding some errors in the key~\cite{KGR05}. 

Our results benefitted from discussions with Daniel Gottesman and Hoi-Kwong Lo, whose contributions are greatly appreciated. We thank Nicolas Gisin who proposed using reverse reconciliation. We also thank Tony Anderson for his assistance. J.-C.B. and R.L. acknowledge support from the Government of Ontario, J.B. and R.L. from NSERC, and R.L. from CIAR, MITACS and ARDA.

\end{document}